\documentclass[aapm,mph,preprint,graphicx,showkeys]{revtex4-1}

\usepackage{lineno,hyperref}
\usepackage{graphicx}
\modulolinenumbers[5]

\begin{document}
\title{Commissioning of a clinical pencil beam scanning proton therapy unit for ultrahigh dose rates (FLASH)}

\author{Konrad~P.~Nesteruk}
\email{konrad.nesteruk@gmail.com}
\author{Michele Togno}
\author{Martin Grossmann}
\affiliation{Paul Scherrer Institute, Center for Proton Therapy, Villigen, Switzerland}

\author{Anthony~J.~Lomax}
\affiliation{Paul Scherrer Institute, Center for Proton Therapy, Villigen, Switzerland \\
ETH Zurich, Department of Physics, Villigen, Switzerland}
\author{Damien~C.~Weber}
\affiliation{Paul Scherrer Institute, Center for Proton Therapy, Villigen, Switzerland \\
University Hospital Zurich, Department of Radiation Oncology, Switzerland\\
University Hospital Bern, Department of Radiation Oncology, Switzerland}

\author{Jacobus~M.~Schippers}
\affiliation{Paul Scherrer Institute, Division Large Research Facilities, Villigen, Switzerland}

\author{Sairos Safai}
\author{David Meer}
\author{Serena Psoroulas}
\affiliation{ Paul Scherrer Institute, Center for Proton Therapy, Villigen, Switzerland }

\begin{abstract}

{\noindent}\textbf{Purpose:} The purpose of this work was to provide a flexible platform for FLASH research with protons by adapting a former clinical pencil beam scanning gantry to irradiations with ultrahigh dose rates. \\
\textbf{Methods:} PSI Gantry 1 treated patients until December 2018. We optimized the beamline parameters to transport the 250 MeV beam extracted from the PSI COMET accelerator to the treatment room, maximizing the transmission of beam intensity to the sample. We characterized a dose monitor on the gantry to ensure good control of the dose, delivered in spot-scanning mode. We characterized the beam for different dose rates and field sizes for transmission irradiations. We explored scanning possibilities in order to enable conformal irradiations or transmission irradiations of large targets (with transverse scanning). \\
\textbf{Results:} We achieved a transmission of 86~\% from the cyclotron to the treatment room. We reached a peak dose rate of 9000 Gy/s at 3~mm water equivalent depth, along the central axis of a single pencil beam. Field sizes of up to 5x5~mm$^{2}$ were achieved for single spot FLASH irradiations. Fast transverse scanning allowed to cover a field of 16x1.2~cm$^{2}$. With the use of a nozzle-mounted range shifter we are able to span depths in water ranging from 19.6 to 37.9~cm. Various dose levels were delivered with a precision within less than 1~\%.
\textbf{Conclusions:} We have realized a proton FLASH irradiation setup able to investigate continuously a wide dose rate spectrum, from 1 to 9000 Gy/s in a single spot irradiation as well as in the pencil beam scanning mode. As such, we have developed a versatile test bench for FLASH research.

\end{abstract}

\keywords{Ultrahigh dose rates, FLASH, proton therapy, gantry, pencil beam scanning}

\maketitle 

\section{Introduction}
One of the limiting factors in delivering radiation therapy to particularly aggressive or radiation resistant tumors is the limitation on the amount of dose healthy tissues can withstand. In recent years, several studies have indicated that ultrahigh dose rates might result in reduced toxicities to healthy tissues while keeping the same tumor control as for treatments with standard dose rate levels. In 2014, the so called FLASH effect has been demonstrated for the first time by irradiating lung tumors in mice with 4.5~MeV electrons~\cite{Favaudon2014}. A reduced pulmonary fibrosis was observed with electrons for a dose rate of 60~Gy/s when compared to a conventional dose rate of 0.03 Gy/s. Whole brain irradiation of mice with 4.5~MeV electrons showed a unique memory sparing for dose rates above 100~Gy/s~\cite{Montay2017}. Almost no skin toxicity has been observed in mini-pigs irradiated with electrons, as well as in cat patients with locally advanced squamous cell carcinoma of the nasal planum~\cite{Vozenin2019}. Eventually, in Switzerland, the first ever FLASH treatment of a human patient has been performed in 2018 at the Lausanne University Hospital (CHUV) on a prototype linac for FLASH radiotherapy with electrons~\cite{Bourhis2019}.

Although the majority of the promising studies of the FLASH effects was performed with electrons,  it is conceptually possible that this FLASH effect could be observed with proton irradiation. Several centers in the world, as well as main vendors in the field, started to investigate the possibility of reaching FLASH dose rates with existing treatment units~\cite{Patriarca2018,Beyreuther2019,Buonanno2019,Darafsheh2020,Diffenderfer2020}. Based on the studies with photons and electrons, it has been concluded that the threshold for the FLASH effect is at least 40~Gy/s, at least a factor 10 higher than conventional dose rates. However, the definition of the dose rate is prone to ambiguities, as the beam has its microstructure and the average dose rate may be substantially different from the maximum dose rate in a short pulse. The definition of the dose rate is also difficult for pencil beam scanning, due to the sequential character of the delivery technique. Since the mechanism of FLASH has not been fully explained, a potential influence of pauses in dose delivery and fractionation is not understood yet. Therefore, exact conditions for the FLASH effect to occur remain unknown, which reflects in non-consistent results of studies performed with protons so far. Thus, flexible FLASH test benches, able to provide different dose rates and irradiation conditions, are required.

In this paper, we report on the commissioning of the former clinical unit Gantry 1 at the Paul Scherrer Institute for FLASH research with protons. PSI Gantry 1 was the first gantry in the world to deliver routinely pencil beam scanning (PBS) to cancer patients. The objective of this work was to transform the in-house developed gantry into a versatile platform for FLASH effect studies. As such, the modified gantry will be able to provide various dose rates and allow for single spot irradiations as well as pencil beam scanning with ultrahigh instantaneous dose rates for biological samples and potentially patients. Dose delivery in both transmission and conformal (Bragg peak) modes will be possible. 

\section{Materials and Methods}
\subsection{Pencil beam scanning Gantry 1}
\begin{figure}[t]
    \centering
    \includegraphics[width=0.7\textwidth]{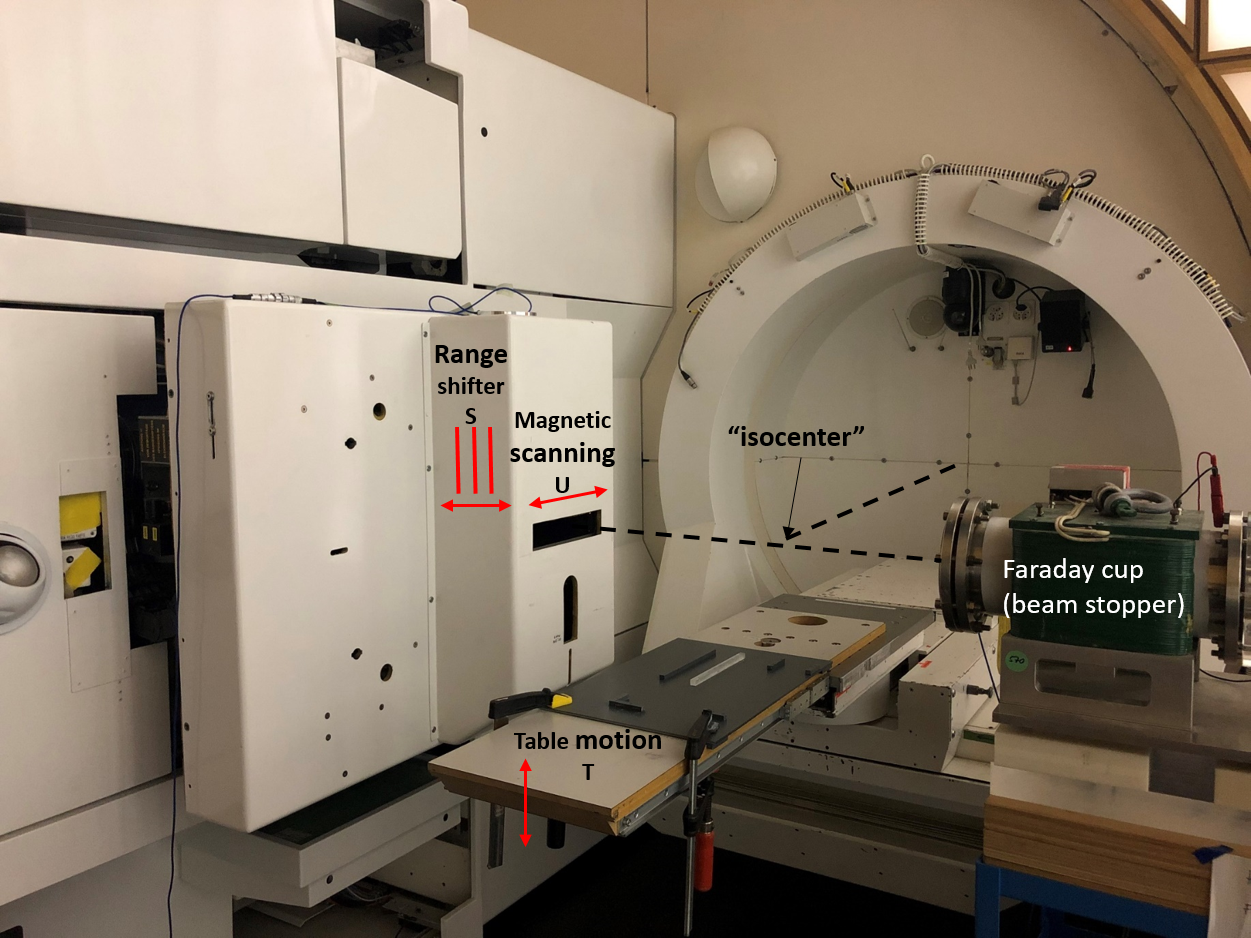}
    \caption{Gantry 1 at an angle of -90 degrees seen from the treatment room. Highlighted are spot-scanning axes and motions, the ''isocenter'' defined for this paper, and a Faraday cup used for experiments.}
    \label{fig:gantry1}
\end{figure}
Gantry 1 treated patients from 1996 until the end of 2018. It was the first facility worldwide that used the spot-scanning technique~\cite{Pedroni1995,Lin2009}. With its radius of 2~m, it is still the most compact gantry in the world competing advantageously with low footprint gantries sold by vendors. The compactness was achieved by mounting the patient table off-axis (excentric gantry) as well as by placing a scanning magnet upstream the last dipole. Since the gantry is excentric and the table moves vertically together with the gantry rotation, there is no true isocenter. In this paper, whenever we refer to the isocenter, we have in mind a moving isocenter defined as a rotation-invariant distance between the nozzle and the axis of the table at its reference position (Fig.~\ref{fig:gantry1}). 

Magnetic scanning is realized in one transverse direction (\textit{U}) by means of a fast scanning magnet, so-called sweeper, with a dead time of a few ms between spots. Scanning in depth (\textit{S}) is achieved with a range shifter in the gantry nozzle, which consists of 39 polystyrene (PS) plates of 4.53~mm thickness and one PS plate of half thickness. They are moved pneumatically in and out of the beam with a dead time of 60 ms per plate. In view of the FLASH applications, this feature gives a unique opportunity to transport an un-degraded proton beam to the isocenter in order to maximize the beam intensity, hence the dose rate, while keeping the possibility of conformal (Bragg peak) irradiations. The slowest scanning direction (\textit{T}) is performed by moving a treatment table. An overview of the gantry, as seen from the treatment room, is presented in Fig.~\ref{fig:gantry1}.

\begin{figure}[t]
    \centering
    \includegraphics[width=0.7\textwidth]{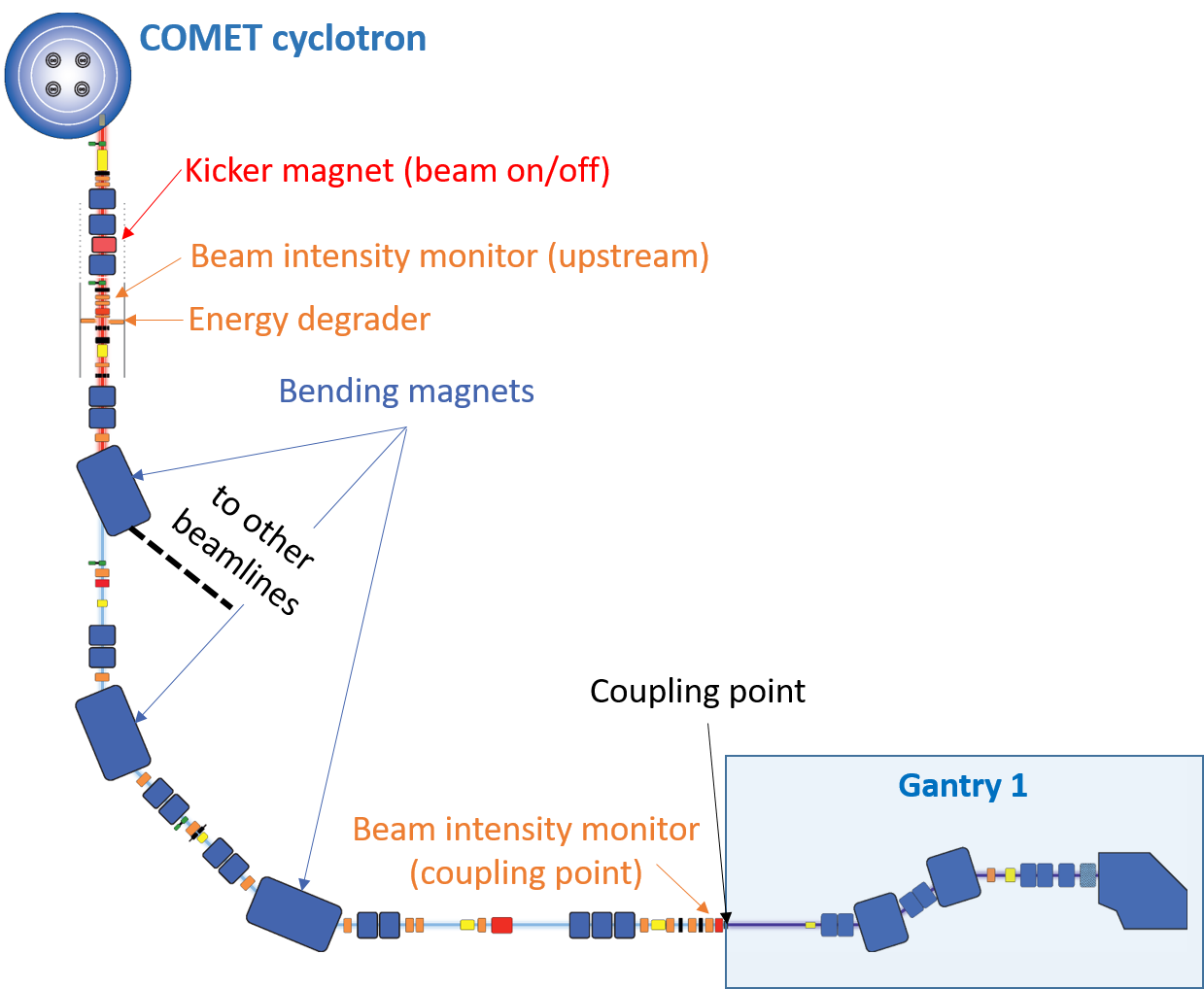}
    \caption{Schematic diagram of the Gantry 1 beamline. Selected components relevant to this paper are highlighted.}
    \label{fig:proscanG1}
\end{figure}

\subsection{Beam transport for ultrahigh dose rates}
Proton beam used for treatments is produced by the COMET cyclotron~\cite{Schippers2007}. The energy and intensity of the extracted beam are 250~MeV and up to 1~$\mu$A, respectively. Energy modulation is performed with a degrader system, installed right after the cyclotron. In the case of Gantry 1 the previously mentioned range shifter was used for an additional fine modulation of the energy pre-selected by the degrader. However, the degrader introduces massive intensity losses due to the scattering occuring in the degrader material. At low energies, the fraction of the beam transported to the treatment room is lower than 1~\%. In order to reach ultrahigh dose rates, we had to optimize the beamline parameters to transport the 250~MeV beam produced by the PSI COMET cyclotron to the treatment room with minimum losses. This energy has never been used for treatments, although the gantry was design to accept energies up to 300~MeV. Therefore, a completely new configuration of beamline magnet settings, so-called beamline tune, had to be found.

The beamline (Fig.~\ref{fig:proscanG1}), including the gantry, is 44~m long and consists of about 40 magnets - dipoles and quadrupoles. Beam transport can be simulated in numerous codes for beam optics calculations. We used a simulation by means of TRANSPORT~\cite{transport1,transport2} to find initial magnet settings and as a guide for further experimental fine tuning. We considered the following criteria for the desired beam optics:
\begin{itemize}
    \item beam size (2-sigma) always much smaller than apertures of magnets and collimators
    \item point-to-point imaging between the location of the degrader (upstream) and the gantry coupling point
    \item limited beam tilt at the entrance of the gantry
    \item beam waist in the gantry at the same location in both bending (dispersive) and non-bending planes
    \item limited beam divergence of the beam extracted from the gantry 
    \item round beam spot at a certain distance from the gantry
    \item minimized dispersion and its derivative at the isocenter.
\end{itemize}

The experimental tuning was based on numerous beam profilers and intensity monitors located along the beamline. It was realized in two stages. The goal of the first stage was to transport the beam to the gantry coupling point with minimum loses. At the second stage, we optimized the transport through the gantry to achieve the maximum transmission, a round beam spot at a defined distance from the gantry nozzle, and a limited beam divergence in air. For these optimizations, we used a scintillating foil and a in-room camera for online tuning, and a scintillator-CCD detector for precise measurements of the beam size and its envelope in air. The beam transmission to the gantry coupling point was determined as a ratio between coupling-point and cyclotron-extracted beam intensities, measured simultaneously by means of intensity monitors installed along the beamline. The beam current at the isocenter was calculated based on the total proton charge measured by a Farady cup (see Fig.~\ref{fig:gantry1}) and measured delivery time. The transmission from the coupling point to the isocenter was defined as a ratio between the calculated Faraday cup current and the current measured at the coupling point.

\subsection{Dose monitoring in the FLASH regime}
Gantry 1 is equipped with two parallel-plane ionization chambers placed in the nozzle to control the dose to be delivered~\cite{Lin2009}. The first chamber (monitor 1) is the main monitor which controls dose delivery with an ion collection time of 90~$\mu$s, while the second chamber (monitor 2) with a collection time of 350~$\mu$s is a backup element and it was part of the therapy verification system. Assuming a FLASH irradiation time of the order of 10~ms, monitor 1 guarantees a precision of at least 1~\%.

Compared to the clinical setting, the instantaneous dose rate we expect in FLASH experiments is 100 to 1000 times higher. Ionization chambers at such high dose rates are affected by significant ion recombination losses and therefore, a thorough characterization of these kind of monitors is needed~\cite{dosimetry_preprint}. In order to provide an accurate dose monitoring, we have performed such a calibration of the nozzle monitor against a Faraday cup (see Fig.~\ref{fig:gantry1}). We used the same Faraday cup to validate the precision of the dose delivery in terms of reproducibility. The prediction of the absolute dose with different dose rates was based on a model described in~\cite{dosimetry_preprint}.

\subsection{Delivery of different dose rates}
For the definition of dose rate we assume the beam produced by our cyclotron is continuous, as the RF frequency of the COMET cyclotron is 72.85 MHz. As such, every 14 ns a 0.8 ns long pulse is delivered. Therefore, in a single pulse the dose rate is 17.5 times higher than under the assumption of continuous beam. However, a single pulse is so short that the dose delivered within it is extremely low. Even for a FLASH dose rate of 1000~Gy/s, the dose per pulse is of the order of 10~$\mu$Gy. Also in terms of ion recombination, the dose delivery can be considered continuous~\cite{Goma2014}.

\subsubsection{Single spot irradiations}
\begin{figure}[thb]
    \centering
    \includegraphics[width=0.7\textwidth]{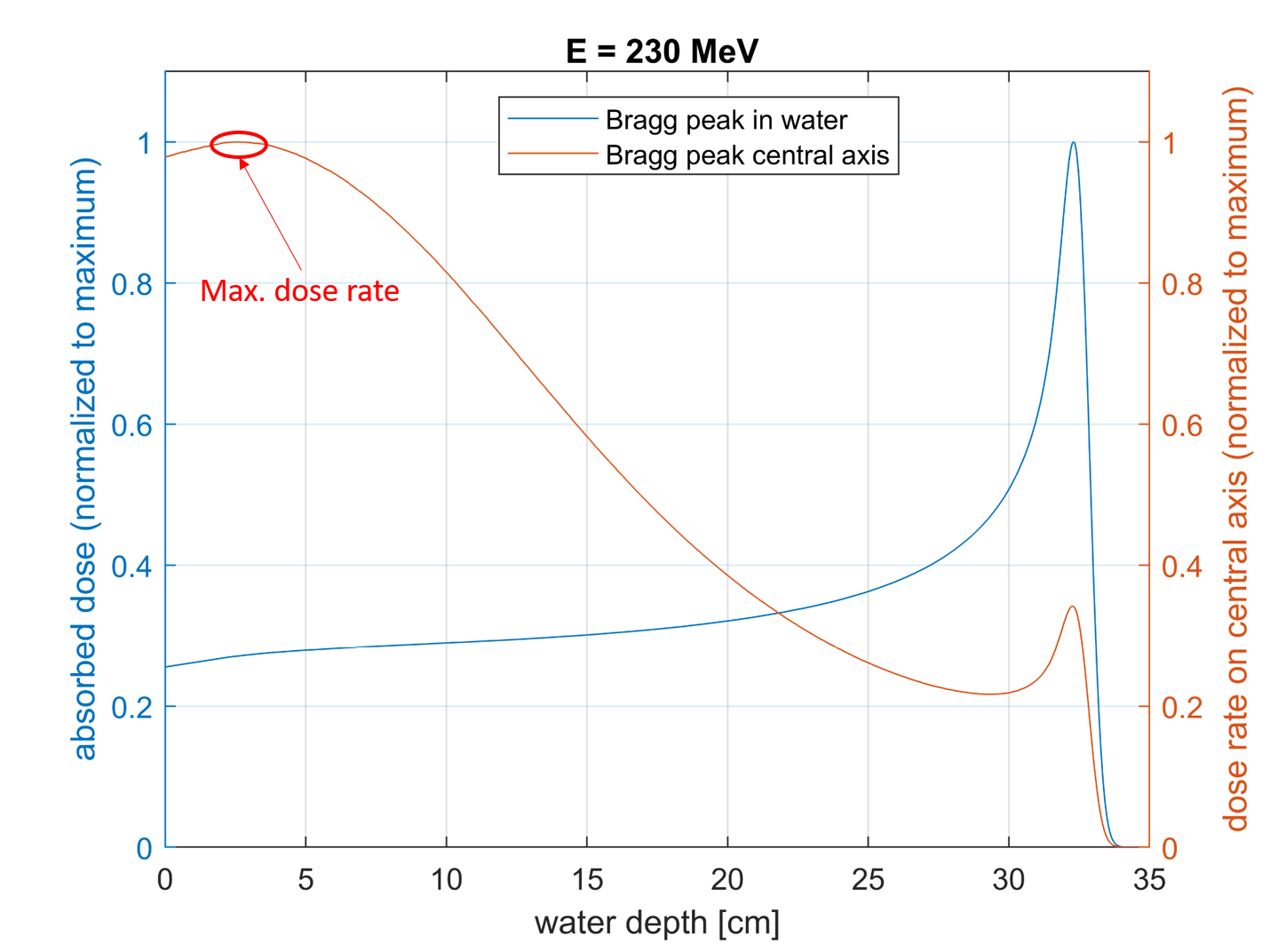}
    \caption{Comparison of Bragg peak curves - integral and on the central axis for a beam energy of 230~MeV. The flat part of the curve with the maximum dose rate is highlighted.}
    \label{fig:braggpeak}
\end{figure}

Single spot transmission irradiations are used to deliver homogeneously a dose to small targets, such as cell lines. In order to obtain homogeneity in depth we use the flat part of the Bragg peak curve on the central axis where the dose, the beam size, and hence the dose rate, is typically constant for a few centimeters, as depicted in Fig.~\ref{fig:braggpeak} and described in~\cite{dosimetry_preprint}. For the transverse homogeneity, we define the field size to be the 95~\% isodose surface for a given Gaussian pencil beam.

The instantaneous dose rate can be varied by adjusting the beam current and beam size. The former is realized by a vertical deflector in the COMET cyclotron. As such, we can vary the cyclotron current between 0.1 and 800 nA. Up to 20~\% higher beam currents are technically achievable. However, they are not clinically commissioned and thus were not used in this study. The variation of beam size can be achieved by choosing a different number of range shifter plates in the gantry nozzle. 
Due to multiple Coulomb scattering the beam size increases and hence the dose rate decreases in a quadratic manner. This can also be used to increase the field size with 95~\% homogeneity at the cost of decreased dose rate.

In order to characterize the beam and estimate the corresponding dose rate we measured beam sizes by means of a CCD-scintillator detector and depth-dose curves in water by means of a large sensitive area, plane-parallel Bragg peak ionization chamber. The dose rates were determined from the measured delivered protons by means of a Faraday cup and recorded delivery time. The time is recorded by the control system and corrected offline for the reaction time of an upstream kicker magnet which switches the beam off when the requested dose is reached.  As such, an accuracy better than 50~$\mu$s is achieved. 

\subsubsection{Pencil beam scanning}
We explored scanning possibilities in order to enable conformal irradiations or transmission irradiations of large targets (transverse scanning).

Since the upstream degrader is not used anymore, the only possibility to scan in depth is to use the range shifter in the gantry nozzle. We verified the minimum reachable energy with all the range shifter plates inserted by performing a range measurement in water.

For scanning in the transverse \textit{U} direction (the bending direction), we evaluated the performance of the scanning magnet (sweeper) at an energy of 250~MeV, as this energy has never been used for treatments. For the main FLASH-optimized tune, a new scanning magnet calibration, so-called sweeper map, was defined to obtain equal spacing between spots. Additionally, we explored the possibility of fast scanning in the \textit{T} direction (the non-dispersive direction) with the use of a steering magnet usually utilized for the first spot correction (beam centering). This magnet is not as fast as the sweeper. With a dead time between consecutive spots of the order of 100~ms it would, however, guarantee much faster scanning in \textit{T} than the table moving. Eventually, a 2D scanning map has been determined. For all the transverse scanning measurements, a CCD-scintillator detector was used.      

\section{Results}
\subsection{FLASH-optimized beam transport}
\begin{figure}[b]
    \centering
    \includegraphics[width=0.7\textwidth]{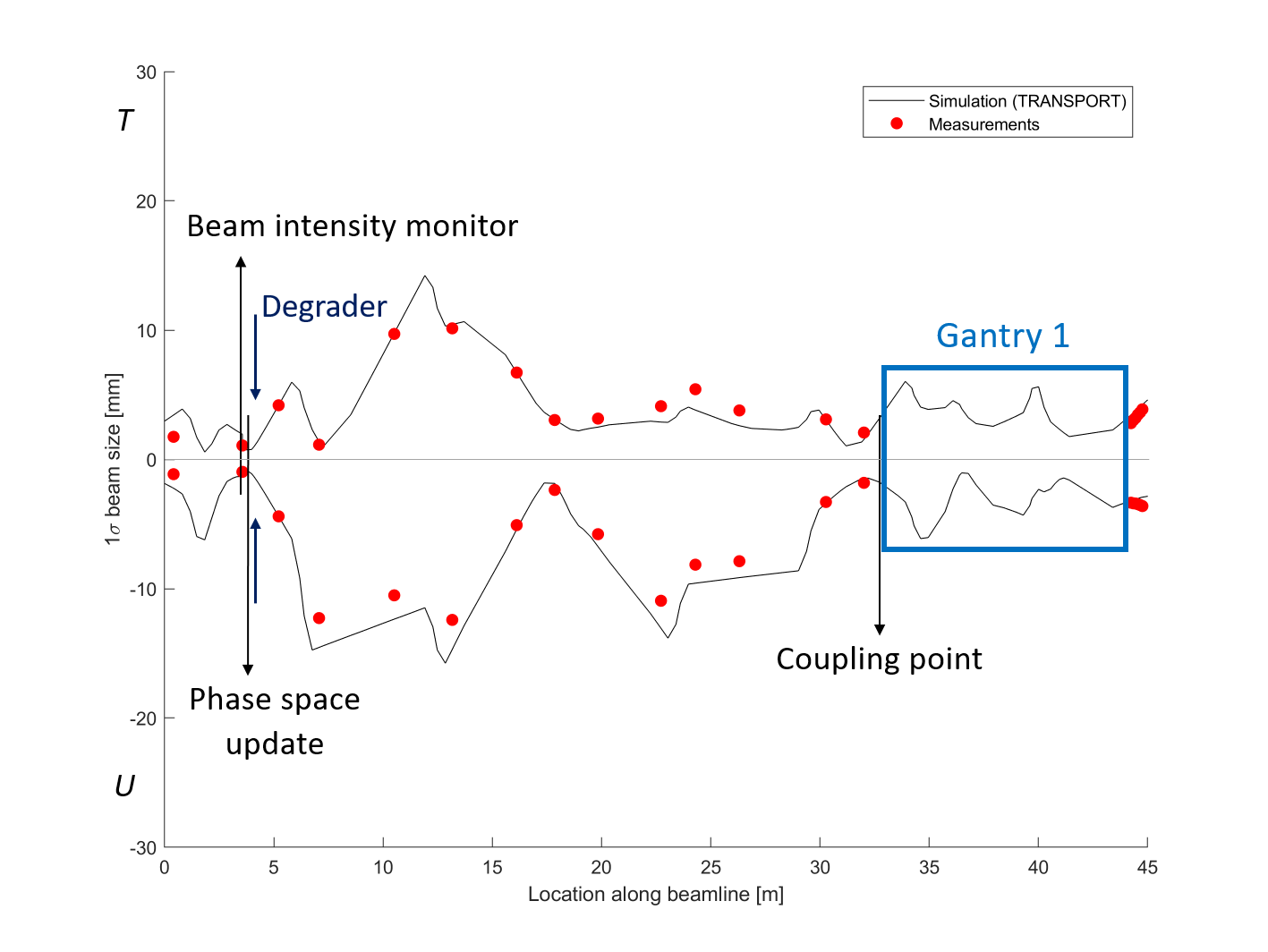}
    \caption{Beam envelope corresponding to the main FLASH beamline tune. Negative values of beam size correspond to the dispersive plane (\textit{U}) and positive values to the non-dispersive plane (\textit{T}). Continuous line - result of simulations with TRANSPORT; red dots - beam profile measurements.}
    \label{fig:tune}
\end{figure}

We characterized two different beamline tunes to be used for FLASH experiments. The main FLASH-optimized tune is shown in Fig.~\ref{fig:tune}. The measured beam envelope is compared with the one simulated in TRANSPORT with the final magnet settings. Due to the design of the compact gantry, there is no beam profile monitor in the gantry. The observed discrepancies are mostly due to the phase space definition. The TRANSPORT simulation is based on the phase space measurements performed in 2006. We updated the phase space (highlighted in the figure) on the basis of upstream beam profile measurements. However, the beam emittance was assumed to be the same as measured in 2006. The agreement we reached is satisfactory for our purpose and the simulation in TRANSPORT served as a useful guide for the beamline tuning, providing an important qualitative description of our beam transport.  The tune presented in Fig.~\ref{fig:tune} provides the highest transmission to the isocenter and the minimum eccentricy of the beam spot, even when the range shifter is not used. The beam divergence in air was measured to be ($0.5\pm0.3$)~mrad and ($2.0\pm0.3$)~mrad for the dispersive (\textit{U}) and non-dispersive (\textit{T}) plane, respectively. At a distance of 41~cm from the nozzle, 5~cm behind the isocenter, the beam spot was found to be perfectly round with a 1$\sigma$ size of ($3.5\pm0.1$)~mm.  
The tune reproducibility has been assessed based on several measurements of beam profiles in different days. The measured beam sizes along the beamline were found to be in agreement within 5~\% and those in air within 3~\%. 

The transmission to the gantry coupling point was measured to be ($86\pm1$)~\% for cyclotron currents exceeding 500~nA. A weak dependence on the beam current was observed, as different settings of the vertical deflector slightly affect the initial phase space and hence the beam transport. The minimum transmission was measured to be ($82\pm1$)~\% for the lowest cyclotron currents, 1~nA and below. The transmission from the coupling point to the isocenter was determined to be 100~\% and independent of the beam current.

\subsection{Ultrahigh dose rates}
\subsubsection{Single spot transmission irradiations}
With the main FLASH tune we have achieved field sizes ranging from 2.3x2.3~mm$^{2}$ for a beam energy of 250~MeV to 5x5~mm$^{2}$ for a beam energy of 170~MeV (all range shifter plates inserted). The corresponding maximum peak dose rates are $(3600\pm200)$~Gy/s and $(700\pm30)$~Gy/s, respectively. The mentioned peak dose rates correspond to 3.5~cm water equivalent depth and a cyclotron beam current of 800~nA. For a given field size, we are able to continuously vary the dose rate from the level of the conventional PBS dose rate (order of 1~Gy/s) to the maximum dose rate corresponding to the given field size.

\begin{figure}[t]
    \centering
    \includegraphics[width=0.7\textwidth]{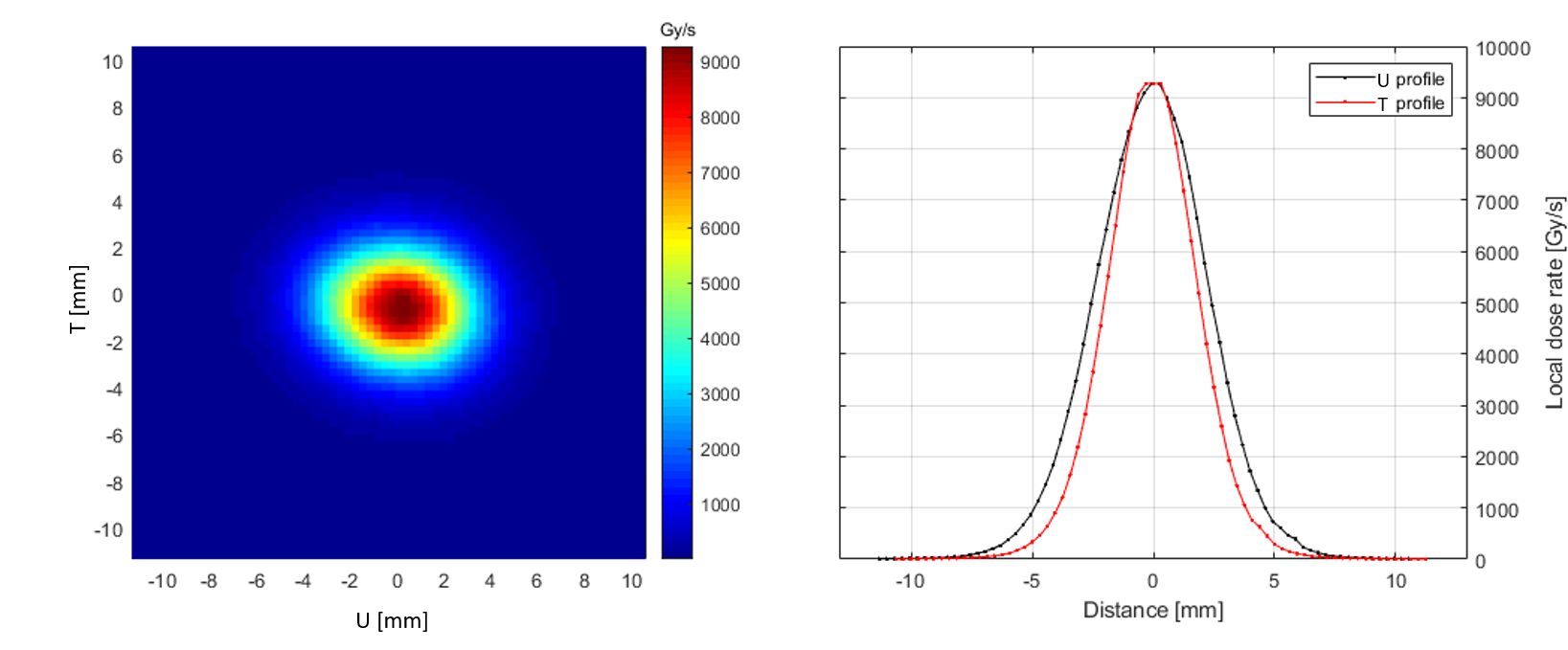}
    \caption{Characterization of a beam spot in air. Left: Beam spot of 2.3/1.8~mm (\textit{U}/\textit{T} - 1$\sigma$) in air recorded with a scintillator-CCD detector.  Right: Corresponding dose rate distribution in water at 3~mm water equivalent depth.}
    \label{fig:doserate}
\end{figure}

The absolute maximum we have achieved was $(9300\pm500)$~Gy/s at 3~mm water equivalent depth. This dose rate was reached with another FLASH tune, which provided a narrower pencil beam. However, for this tune the beam spot is not round, when no range shifter plate is used. At a distance of 56~cm from the gantry nozzle, 10~cm behind the isocenter, 1$\sigma$ sizes of the beam spot were measured to be ($2.27\pm0.05$)~mm and ($1.80\pm0.05$)~mm in the~\textit{U} and~\textit{T} directions, respectively. As such, the maximum field size with this beam spot for transmission irradiations is 1.1x1.1~mm$^{2}$. The characterization of the beam spot in air for this maximum dose rate is depicted in Fig.~\ref{fig:doserate}.

The uncertainty of the dose rate consists of the model uncertainty, dose delivery uncertainty and the beam size uncertainty. The beam size uncertainty contributes most to the total uncertainty. In our characterization studies, the uncertainty due to the delivery time measurement was negligible due to 3 to 4 orders of magnitude longer delivery times than the precision of the time measurement.

\begin{figure}[tbh]
    \centering
    \includegraphics[width=0.7\textwidth]{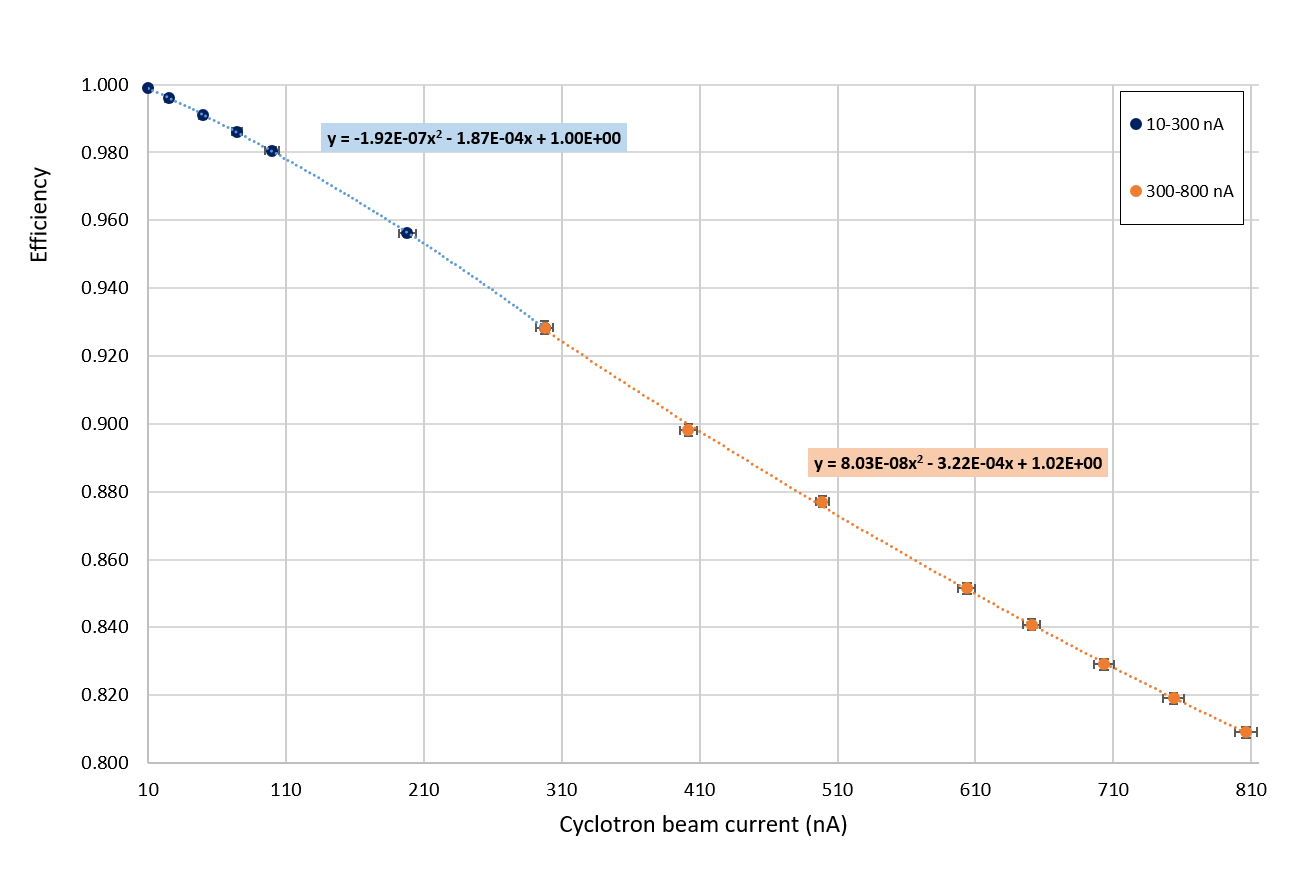}
    \caption{Efficiency of monitor~1 in the gantry nozzle as a function of cyclotron beam current.}
    \label{fig:monitor1}
\end{figure}

In order to deliver different dose levels with a wide range of dose rates we used the previously described monitor~1 calibrated against dose rate independent Faraday cup. For all cyclotron beam currents up to 5~nA we did not observe any efficiency drop of monitor~1 due to ion recombination and the ratio between monitor units (MU) of monitor~1 and the total charge $q$ measured with the Faraday cup remained constant. Also between 5 and 10~nA the efficiency drop was very subtle, always below 1~\%. For cyclotron currents exceeding 10~nA, the monitor efficiency, normalized to the MU$/Q$ ratio corresponding to a cyclotron current of 1~nA, was observed to drop rapidly with the increasing current, as shown in Fig.~\ref{fig:monitor1}. The corresponding calibration curve had to be divided in two regions - moderate current range (10-300~nA) and high current range (300-800~nA), as the monitor response presents a kink at about 300~nA and another polynomial had to be fitted to the latter cyclotron current range. With this calibration we were able to deliver different doses with different dose rates with a precision always better than 1~\%, as verified with the Faraday cup.

\subsubsection{Pencil beam scanning mode}
\begin{figure}[tbh]
    \centering
    \includegraphics[width=0.7\textwidth]{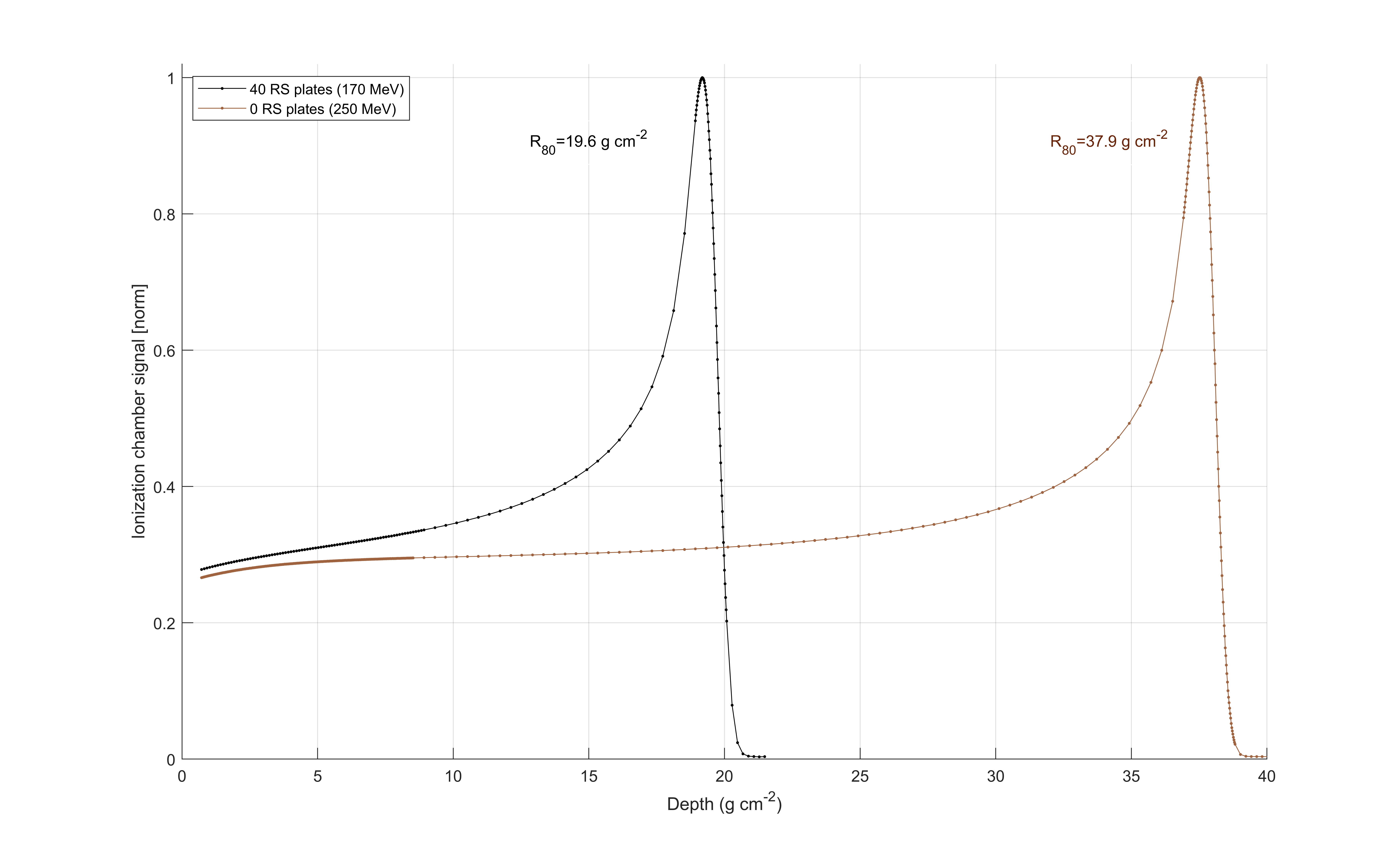}
    \caption{Bragg peak curves measured with a range scanner for two minimum and maximum beam energies.}
    \label{fig:range}
\end{figure}

The range shifter enables scanning in depth between 19.6~g$\cdot$cm$^{-2}$ and 37.9~g$\cdot$cm$^{-2}$ (R$_{80}$ ranges of the proton beam at 80~\% distal fall-off). The measured ranges (Fig.~\ref{fig:range}) correspond to the energies 250~MeV and 170~MeV, respectively. The former is achieved when no range shifter (RS) plate is used, the latter corresponds to the use of all the 40 RS plates.

\begin{figure}[tbh]
    \centering
    \includegraphics[width=0.7\textwidth]{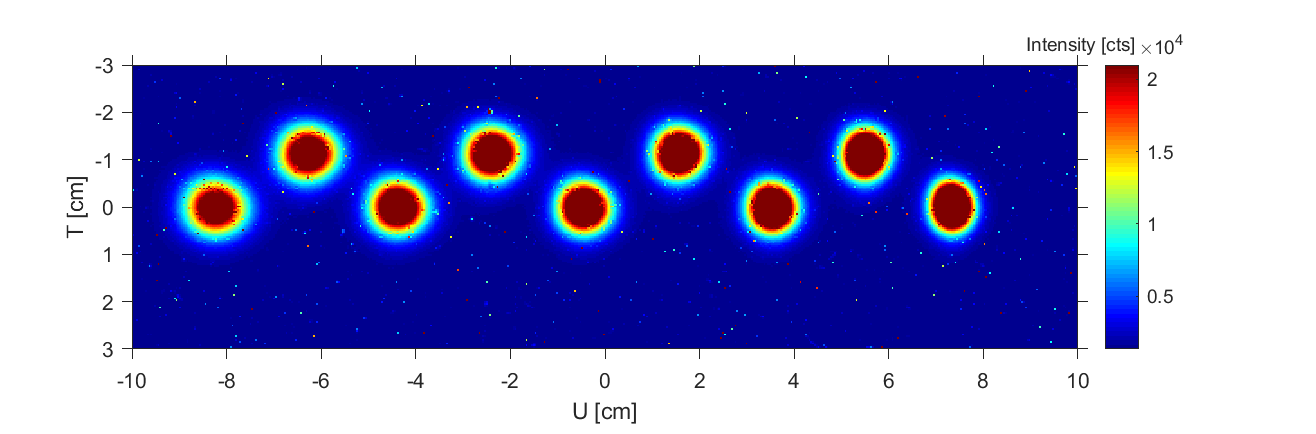}
    \caption{Transverse pencil beam scanning pattern acquired with a CCD-scintillator detector.}
    \label{fig:scanning}
\end{figure}

The maximum range of the transverse scanning with the minimum beam spot distortion was found to be 16~cm ($\pm8$~cm) and 1.2~cm ($+0.5$/$-0.7$~cm) in the~\textit{U} and \textit{T} directions, respectively. In Fig.~\ref{fig:scanning} a scanning pattern acquired with a CCD-scintillator detector is shown. The pattern is based on a previously measured 2D scanning map. The spots are separated by 2~cm in the~\textit{U} direction and by 1.2~cm (maximum range) in the ~\textit{T} direction.

\section{Discussion}
In this paper, we report the commissioning of the former treatment unit Gantry 1 at PSI for FLASH experiments. In order to maximize the dose rate in the treatment room, we optimized the beam transport for 250~MeV beam, and decided to use only the in-room range shifter for the energy degradation. We achieved a transmission of $86\%$, i.e. we could reach more than 680 nA beam irradiation at isocenter. Higher beam intensity could be potentially be achieved, as already shown in a proof-of-principle study by Busold~\textit{et al}~\cite{Busold2017}. Moreover, losses have been observed only upstream of the treatment room, and no loss has been observed on the gantry, therefore satisfying the radiation protection requirements of our facility (where the shielding of the treatment room is thinner than the one of the cyclotron bunker, which can sustain higher beam losses). In order to minimize the losses we refrained from using collimators which help define better the phase space and make the beam transport less sensitive to cyclotron phase space variations. This contributes to our beam size uncertainty and thus the total uncertainty of the dose rate.   

Such high beam currents can then be shaped to a dose and dose-rate distribution using either multiple- or single-spot deliveries. The latter provides the highest instantaneous dose rates available at our facility: we could achieve more than 9000 Gy/s, which matches dose-rate levels achieved at synchrotron facilities such as the European Synchrotron Research Facility (ESRF) - 8000~Gy/s and 16000~Gy/s~\cite{Serduc2006,Braeuer-Krisch2015} - and electron machines like the eRT6 electron linac at Lausanne University Hospital (CHUV). The gantry control system does not limit the length of these single-spot irradiations, therefore we can deliver all possible doses in both single- and multiple-spots deliveries. The minimum pause between two spots corresponds to 2~ms, but larger pauses can be introduced by the control system, to help investigating the impact of these variables on irradiation modalities such as clinical spot scanning. This minimum time can be further decreased down to 100~$\mu$s by using the vertical deflector in the cyclotron and following some modifications of the control system.

The dose can be precisely delivered based on the nozzle monitor, which we calibrated against the Faraday cup. However, we observed some day-to-day instabilities in the calibration. Therefore, we apply a day-specific correction for our calibration by measuring the monitor-1-to-Faraday-cup ratio (MU$/Q$) for low and high dose rates, which takes a few minutes. We observed an increase ratio over time since the original calibration in the MU$/Q$ ratio. We have not understood this effect yet. One of the explanations could be temperature and pressure changes, for which we do not apply any correction. The maximum deviations we have observed so far were up to 4~\% and can be completely neutralized by the mentioned day-specific correction.  

Even though the energy can be modified easily inside the treatment room, the presence of the range shifter at the end of the beamline causes a clear correlation between final energy and dose rate. The beam size increases as function of number of range shifter plates and distance from the gantry exit, and consequently the dose rate drops. At the minimum energy achievable with our system, 700 Gy/s could be reached at the isocenter. Such performance could be improved in case a different material for the range shifters is used, or in case the samples are placed closer to the gantry exit.

On the other hand, the relationship between range shifters and beam size allows an easy way of adapting the beam conditions to the size of the sample to irradiate, without the production of complex single or double-scattering systems. We have tested different combinations of number of range shifter plates and distances from the gantry nozzle, simulating different sample sizes. This makes our experimental setup particularly interesting for experiments with cells, as we could adapt flexibly to the requirements of external users. 

Thanks to its flexibility, PSI Gantry 1 will give a chance to address a few important questions concerning the FLASH effect mechanism and to define appropriate dose delivery conditions. Different studies of the FLASH effect showed different results, from no protective effect of FLASH irradiations to a significant reduction of toxicity to healthy tissues. However, there are several aspects which have to be taken into account while comparing various results obtained with different beam parameters. One of the most important is the difference between the average and instantaneous dose rate. Depending on the machine type and dose delivery technique, the two dose rates may be very different. For instance, the Kinetron and Oriatron eRT6 electron linacs, used at Orsay and Lausanne University Hospital (CHUV), respectively, allow an instantaneous dose rate in a 1-2~$\mu$s pulse to exceed 10$^{7}$~Gy/s~\cite{Favaudon2014,Lansonneur2019,Jaccard2018,Petersson2017}. However, the average dose rate under typical treatment conditions would be ''only'' up to a few~kGy/s and would involve pauses between pulses of at least 5~ms.  With these machines the most promising results have been achieved so far. However, if the instantaneous dose rate plays the key role in triggering of the FLASH effect, it could well be that reaching an average dose rate of the order of 100~Gy/s or even 1000~Gy/s will not be sufficient unless large enough instantaneous dose rate is provided. On the other hand, studies in different animal and cells models have shown FLASH effects at such average dose rates. Dose-rate thresholds observed are not fully consistent among different studies, though, and this prompted some authors to suggest that other factors might play a role - not only the instantaneous dose rate, but also the total irradiation time as well as the minimum dose within a (macro-)pulse~\cite{Bourhis2019b}. Different machines are currently able to test at best only a part of the parameter space, as highlighted in a recent overview of the so-called FLASH-compatible machines~\cite{Esplen2020}; therefore it is of fundamental importance to develop flexible irradiation facilities, allowing to test in the same conditions and different animal models different combinations of dose, dose rate and pulse structure. 


\section{Conclusions and outlook}
We have successfully converted the PSI pencil beam scanning Gantry 1 into a test bench for FLASH experiments. We are able to deliver a wide range of dose rates from 1 to 9000~Gy/s, which enables detailed studies of FLASH with protons. Moreover, we can conduct transmission as well as conformal irradiations using only local energy modulation. Field sizes larger than 5x5~mm$^{2}$ can be irradiated by means of fast transverse pencil beam scanning. Although the gantry has not been designed to provide scanning in both transverse directions, we have proven that it is possible to include the second direction for fast scanning of small fields. This feature can be used for pre-clinical studies of FLASH with small animals.

Teaming with colleagues from CHUV and Varian Medical Systems, we are currently carrying out our first biological experiments at Gantry 1.

\begin{acknowledgments}
We thank our collaborators at PSI (M. Schippers, M. Eichin, U. Rechsteiner, B. Rohrer, M. Egloff) and CHUV (M.-C. Vozenin, V. Grilj, C. Bailat). This work is partially funded by the Swiss National Science Foundation (grant No. 190663).
\end{acknowledgments}

\section*{Conflict of interest statement}
The authors have no conflict to disclose.

\bibliography{mybibfile}

\begin{thebibliography}{10}

\bibitem{Favaudon2014}
V.~Favaudon, L.~Caplier, V.~Monceau, F.~Pouzoulet, M.~Sayarath, C.~Fouillade,
  M.-F. Poupon, I.~Brito, P.~Hup{\'e}, J.~Bourhis, J.~Hall, J.-J. Fontaine, and
  M.-C. Vozenin,
\newblock Ultrahigh dose-rate FLASH irradiation increases the differential
  response between normal and tumor tissue in mice,
\newblock Science Translational Medicine {\bf 6}, 245ra93 (2014).

\bibitem{Montay2017}
P.~Montay-Gruel, K.~Petersson, M.~Jaccard, G.~Boivin, J.-F. Germond, B.~Petit,
  R.~Doenlen, V.~Favaudon, F.~Bochud, C.~Bailat, J.~Bourhis, and M.-C. Vozenin,
\newblock Irradiation in a flash: Unique sparing of memory in mice after whole
  brain irradiation with dose rates above 100Gy/s,
\newblock Radiotherapy and Oncology {\bf 124}, 365 -- 369 (2017),
\newblock 15th International Wolfsberg Meeting 2017.

\bibitem{Vozenin2019}
M.-C. Vozenin, P.~De~Fornel, K.~Petersson, V.~Favaudon, M.~Jaccard, J.-F.
  Germond, B.~Petit, M.~Burki, G.~Ferrand, D.~Patin, H.~Bouchaab, M.~Ozsahin,
  F.~Bochud, C.~Bailat, P.~Devauchelle, and J.~Bourhis,
\newblock The Advantage of FLASH Radiotherapy Confirmed in Mini-pig and
  Cat-cancer Patients,
\newblock  {\bf 25}, 35--42 (2019).

\bibitem{Bourhis2019}
J.~Bourhis, W.~J. Sozzi, P.~G. Jorge, O.~Gaide, C.~Bailat, F.~Duclos, D.~Patin,
  M.~Ozsahin, F.~Bochud, J.-F. Germond, R.~Moeckli, and M.-C. Vozenin,
\newblock Treatment of a first patient with FLASH-radiotherapy,
\newblock Radiotherapy and Oncology {\bf 139}, 18 -- 22 (2019),
\newblock FLASH radiotherapy International Workshop.

\bibitem{Patriarca2018}
A.~Patriarca, C.~Fouillade, M.~Auger, F.~Martin, F.~Pouzoulet, C.~Nauraye,
  S.~Heinrich, V.~Favaudon, S.~Meyroneinc, R.~Dendale, A.~Mazal, P.~Poortmans,
  P.~Verrelle, and L.~{De Marzi},
\newblock Experimental Set-up for FLASH Proton Irradiation of Small Animals
  Using a Clinical System,
\newblock International Journal of Radiation Oncology*Biology*Physics {\bf
  102}, 619 -- 626 (2018).

\bibitem{Beyreuther2019}
E.~Beyreuther, M.~Brand, S.~Hans, K.~Hideghéty, L.~Karsch, E.~Leßmann,
  M.~Schürer, E.~R. Szabó, and J.~Pawelke,
\newblock Feasibility of proton FLASH effect tested by zebrafish embryo
  irradiation,
\newblock Radiotherapy and Oncology {\bf 139}, 46 -- 50 (2019),
\newblock FLASH radiotherapy International Workshop.

\bibitem{Buonanno2019}
M.~Buonanno, V.~Grilj, and D.~J. Brenner,
\newblock Biological effects in normal cells exposed to FLASH dose rate
  protons,
\newblock Radiotherapy and Oncology {\bf 139}, 51 -- 55 (2019),
\newblock FLASH radiotherapy International Workshop.

\bibitem{Darafsheh2020}
A.~Darafsheh, Y.~Hao, T.~Zwart, M.~Wagner, D.~Catanzano, J.~F. Williamson,
  N.~Knutson, B.~Sun, S.~Mutic, and T.~Zhao,
\newblock Feasibility of proton FLASH irradiation using a synchrocyclotron for
  preclinical studies,
\newblock Medical Physics {\bf 47}, 4348--4355 (2020).

\bibitem{Diffenderfer2020}
E.~S. Diffenderfer et~al.,
\newblock Design, Implementation, and in Vivo Validation of a Novel Proton
  FLASH Radiation Therapy System,
\newblock International Journal of Radiation Oncology*Biology*Physics {\bf
  106}, 440 -- 448 (2020).

\bibitem{Pedroni1995}
E.~Pedroni, R.~Bacher, H.~Blattmann, T.~Böhringer, A.~Coray, A.~Lomax, S.~Lin,
  G.~Munkel, S.~Scheib, U.~Schneider, and A.~Tourovsky,
\newblock The 200-MeV proton therapy project at the Paul Scherrer Institute:
  Conceptual design and practical realization,
\newblock Medical Physics {\bf 22}, 37--53 (1995).

\bibitem{Lin2009}
S.~Lin, T.~Boehringer, A.~Coray, M.~Grossmann, and E.~Pedroni,
\newblock More than 10 years experience of beam monitoring with the Gantry 1
  spot scanning proton therapy facility at PSI,
\newblock Medical Physics {\bf 36}, 5331--5340 (2009).

\bibitem{Schippers2007}
J.~Schippers, R.~D{\"o}lling, J.~Duppich, G.~Goitein, M.~Jermann, A.~Mezger,
  E.~Pedroni, H.~Reist, and V.~Vrankovic,
\newblock The SC cyclotron and beam lines of PSI's new protontherapy facility
  PROSCAN,
\newblock Nuclear Instruments and Methods in Physics Research Section B: Beam
  Interactions with Materials and Atoms {\bf 261}, 773 -- 776 (2007),
\newblock The Application of Accelerators in Research and Industry.

\bibitem{transport1}
K.~L. Brown, D.~C. Carey, F.~C. Iselin, and F.~Rothacker,
\newblock {\em {TRANSPORT: a computer program for designing charged-particle
  beam-transport systems}},
\newblock CERN Yellow Reports: Monographs, CERN, Geneva, 1980,
\newblock Also publ. as SLAC and FERMILAB.

\bibitem{transport2}
U.~Rohrer,
\newblock {G}raphic {T}ransport {F}ramework, 2007.

\bibitem{dosimetry_preprint}
S.~Safai and et~al,
\newblock Dosimetry and biological experiments with proton beams at ultra-high
  dose rates (FLASH): experimental setup and first dosimetry results,
\newblock Physics in Medicine and Biology  (2020),
\newblock in preparation.

\bibitem{Goma2014}
C.~Gom{\`{a}}, S.~Lorentini, D.~Meer, and S.~Safai,
\newblock Proton beam monitor chamber calibration,
\newblock Physics in Medicine and Biology {\bf 59}, 4961--4971 (2014).

\bibitem{Busold2017}
S.~Busold and H.~Röcken,
\newblock Beam Intensity Modulation Capabilities for Varian’s ProBeam®
  Isochronous Cyclotron,
\newblock Proceedings of the 21st Int. Conf. on Cyclotrons and their
  Applications Cyclotrons2016 , THP05 (2017).

\bibitem{Serduc2006}
R.~Serduc, P.~Vérant, J.-C. Vial, R.~Farion, L.~Rocas, C.~Rémy, T.~Fadlallah,
  E.~Brauer, A.~Bravin, J.~Laissue, H.~Blattmann, and B.~van~der Sanden,
\newblock In vivo two-photon microscopy study of short-term effects of
  microbeam irradiation on normal mouse brain microvasculature,
\newblock International journal of radiation oncology, biology, physics {\bf
  64}, 1519—1527 (2006).

\bibitem{Braeuer-Krisch2015}
E.~Bräuer-Krisch et~al.,
\newblock Medical physics aspects of the synchrotron radiation therapies:
  Microbeam radiation therapy (MRT) and synchrotron stereotactic radiotherapy
  (SSRT),
\newblock Physica medica : PM : an international journal devoted to the
  applications of physics to medicine and biology : official journal of the
  Italian Association of Biomedical Physics (AIFB) {\bf 31}, 568—583 (2015).

\bibitem{Lansonneur2019}
P.~Lansonneur, V.~Favaudon, S.~Heinrich, C.~Fouillade, P.~Verrelle, and L.~{De
  Marzi},
\newblock Simulation and experimental validation of a prototype electron beam
  linear accelerator for preclinical studies,
\newblock Physica Medica {\bf 60}, 50 -- 57 (2019).

\bibitem{Jaccard2018}
M.~Jaccard, M.~Durán, K.~Petersson, J.~Germond, P.~Liger, M.-C. Vozenin,
  J.~Bourhis, F.~Bochud, and C.~Bailat,
\newblock High dose-per-pulse electron beam dosimetry: Commissioning of the
  Oriatron eRT6 prototype linear accelerator for preclinical use,
\newblock Medical Physics {\bf 45}, 863--874 (2018).

\bibitem{Petersson2017}
K.~Petersson, M.~Jaccard, J.-F. Germond, T.~Buchillier, F.~Bochud, J.~Bourhis,
  M.-C. Vozenin, and C.~Bailat,
\newblock High dose-per-pulse electron beam dosimetry — A model to correct
  for the ion recombination in the Advanced Markus ionization chamber,
\newblock Medical Physics {\bf 44}, 1157--1167 (2017).

\bibitem{Bourhis2019b}
J.~Bourhis, P.~Montay-Gruel, P.~{Goncalves Jorge}, C.~Bailat, B.~Petit,
  J.~Ollivier, W.~Jeanneret-Sozzi, M.~Ozsahin, F.~Bochud, R.~Moeckli, J.-F.
  Germond, and M.-C. Vozenin,
\newblock Clinical translation of FLASH radiotherapy: Why and how?,
\newblock Radiotherapy and Oncology {\bf 139}, 11 -- 17 (2019).

\bibitem{Esplen2020}
N.~M. Esplen, M.~S. Mendonca, and M.~Bazalova-Carter,
\newblock Physics and biology of ultrahigh dose-rate (FLASH) radiotherapy: a
  topical review,
\newblock Physics in Medicine and Biology  (2020).

\end{thebibliography}

\end{document}